# Measurement and modeling of the mechanical and electrochemical response of amorphous Si thin film electrodes during cyclic lithiation

Giovanna Bucci, Siva P. V. Nadimpalli, Vijay A. Sethuraman, Allan F. Bower, Pradeep R. Guduru,

School of Engineering, Brown University, Providence, Rhode Island 02912, USA

[z]Corresponding author, email: allan_bower@brown.edu, Tel: +1 401 863 1493, Fax: +1 401 863 9009

## Abstract

A combination of experimental measurements and numerical simulations are used to characterize the mechanical and electrochemical response of thin film amorphous Si electrodes during cyclic lithiation. Parameters extracted from the experiment include the variation of elastic modulus and the flow stress as functions of Li concentration; the strain rate sensitivity; the diffusion coefficient for Li transport in the electrode; the free energy of mixing as a function of Li concentration in the electrode; the exchange current density for the Lithium insertion reaction; as well as reaction rates and diffusion coefficients characterizing the rate of formation of solid-electrolyte interphase layer at the electrode surface. Model predictions are compared with experimental measurements; and the implications for practical Si based electrodes are discussed.

## 1. Introduction

Silicon has a charge capacity that is nearly ten times greater than the graphite-based materials that are currently used as the negative electrode in lithium ion batteries [1]. There is consequently great interest in developing electrodes made from either pure silicon, or a silicon based composite, and a variety of potential electrode designs have been explored, including thin films [2] [3] [4] [5] [6], nanowires [7] [8] [9] [10], single particles [11] [12] [12] and composites [13] [14] [15]. These designs have shown promise, and Si based electrodes are starting to see commercial applications. Panasonic Corporation, for instance, announced the development of high-capacity lithium-ion batteries with silicon-alloy anode [16], but Si based batteries continue to suffer significant loss of capacity with charge-discharge cycles, which limits their use in applications such as electric vehicles.

The poor cycle life of Si based electrodes is caused by fracture and mechanical decrepitation, which result in a loss of active electrode material, and are also accompanied by formation of solid-electrolyte-interphase (SEI) layer on the newly exposed crack surfaces. This reaction consumes Li in the battery and increases its internal resistance, which together eventually render the cell unusable. Fracture is a consequence of large the large volume expansion that occurs as Li is inserted into Si (a volumetric strain of 270% occurs as a Si electrode is fully charged [13]. Managing the stresses generated by this volume expansion is the principal challenge involved in developing Si electrodes. Stresses can be minimized by devising electrode architectures that permit the material to expand freely. For example in [17] the authors developed a Si anode nanostructured in arrays of sealed, tubular geometries that exhibits 85% of initial Coulombic efficiencies and capacity-retention of 80% after 50 cycles. In [18] stable capacities of approximately 3200 mAh/g for almost 20 cycles were obtained with Si nanowire electrodes, while in [19] stable capacities of 2000 mAh/g for 50 cycles were achieved by using 100 nm thick amorphous Si (a-Si) film electrodes. Patterned thin film electrodes on substrates [20] are another promising method for mitigating stress and damage in Si electrodes.

The need to design failure resistant electrode architectures and materials has motivated a large number of theoretical and experimental studies that aim to characterize and quantify stress and damage evolution in Si materials during cyclic lithiation. Qualitative measurements of stress in Si electrodes were first made a decade ago, [21], and since then a number of further experimental studies have provided quantitative measurements of the stress state in lithiated silicon [22] [23] [24]. In conjunction with the experiments, a number of theoretical models that aim to characterize the constitutive behavior of insertion electrode materials have been developed [25] [26] [27]. Recent models have focused on the coupling between mechanical behavior and chemistry [28] [29] [30]. In addition, ab-initio computations have been conducted to determine mechanical properties or electrochemical behavior [31] [32] [33] [34]. These studies have provided the theoretical framework and much of the experimental data that will be needed to model and optimize candidate Si based electrode microstructures. For accurate predictions, it is necessary to determine accurate values for the material parameters in constitutive laws of Si. Parameters of interest include mechanical properties such as the elastic modulus and flow stress of Si as functions of Li concentration; transport properties such as the diffusion coefficient of Li through Si; parameters related to the solution chemistry of Li in Si such as the concentration dependence of the activation coefficient; as well as parameters characterizing the electrochemical reactions and side reactions at the Si/electrolyte interface.

There has been great interest in using computations or experiments to determine material parameters for Si electrodes. For example, the elastic properties of both crystalline and amorphous Li–Si alloys were estimated by Shenoy and Johari [31] using first-principles calculations. They predict that the Young's modulus decreases approximately linearly from 90GPa to 20 GPa as Li concentration increases from zero to full capacity. They attribute the elastic softening to the reduced stiffness of Li–Si bonds in comparison to Si-Si bonds. Experimental measurements [22] of biaxial moduli of a Si thin-film electrode reports a decrease from around 70 GPa to about 35 GPa, as the Li fraction increases. Continuum models typically assume a linear variation of modulus with concentration models [35] [25]. Sethuraman and Guduru [36] also observed plastic flow in Si electrodes during lithiation and de-lithiation, concluding that the material has a flow stress of approximately 1 GPa. In addition, diffusion coefficients for Li transport through Si have been measured experimentally, but the measured diffusivities span several orders of magnitude from $10^{-16}$ to $10^{-10} cm^2 s^{-1}$ [37] [38] [39] [40]. Atomic-scale models report values of Li diffusivity of order of $10^{-12} cm^2 s^{-1}$ for a-Si models [28]. Solution thermodynamic parameters such as the free energy of mixing for Li/Li$_x$Si can be extracted from the potential vs. composition curve available in literature from experimental measurements [41] [42] and *ab-initio* DFT calculations [33] [34]. The electrochemical reaction rates governing Li insertion at the electrode surface are usually characterized using a Butler-Volmer relation [43], which includes several parameters that control the exchange current density (the reaction rates at electrochemical equilibrium. Exchange current densities have been measured by means of cyclic voltammograms: for example Chandrasekaran *et al* [44] have suggested that data can be fit by using two different exchange current density values characterizing the reaction rates during lithiation and de-lithiation. The capacity loss to side reactions that lead to SEI formation on the surface of Si electrodes has also been measured experimentally [45]. Several models of SEI formation and capacity loss in Li-ion battery systems have been proposed [46] [47] [48], which generally treat the process using reaction-diffusion equations that characterize the transport of reagents through the SEI layer and the rate of formation of additional SEI at the electrode surface.

In this paper, our goal is to use a systematic combination of experimental measurements and numerical simulations to characterize the response of sputtered amorphous thin film Si electrodes (selecting a-Si for the electrode avoids the amorphous-crystalline phase transition that occurs during the first lithiation cycle of crystalline Si). An experimental apparatus developed by Sethuraman and Guduru [36] was used to measure *in-situ* the variation of voltage and stress in a-Si thin film electrodes in a Li ion half cell (with Li counter-electrode) which were subjected to cyclic lithiation and de-lithiation at fixed surface current density. The experiment is modeled by extending a continuum description of electrochemical reactions, stress, and plastic deformation in Li insertion materials developed by Bower and Guduru [25]. A comparison of experimental measurements and numerical predictions were used to determine mechanical and chemical properties that best characterize the a-Si electrodes. In addition, a series of 'potentiostatic intermittent titration technique experiments,' [49] [50], were conducted, in which the electrodes were subjected to incremental step changes in potential, and the subsequent transient evolution of stress and electric current in the electrode were measured. The tests are used to determine values for several material parameters that govern deformation, electrochemistry and transport in Si electrodes, including the elastic modulus; the flow stress and strain rate sensitivity; the diffusion coefficient for Li transport in the electrode; the free energy of mixing as a function of Li concentration in the electrode; the exchange current density for the Lithium insertion reaction; as well as parameters governing the rate of formation of solid-electrolyte interphase layer at the electrode surface. Model predictions with the best fit parameters are then compared with experimental measurements; and the implications for practical Si based electrodes are discussed.

The remainder of this paper is organized as follows. In the next section, we review the experimental procedure for measuring stress and electrochemical response of a-Si thin film electrodes. We then describe the model and constitutive equations that we use to characterize the materials, and the approach that we used to determine values for material parameters. Section 4 describes the results of both experiments and model, and section 5 contains conclusions.

2. **Experimental measurements**

The experimental procedure used in this study was described in detail in [36] and so will be reviewed briefly here. The apparatus consists of a Li-ion half-cell, with a thin film amorphous Si working electrode and a Li metal foil counter-electrode, as illustrated in Fig. 1. The cell was assembled and tested in an argon-filled glove box maintained at 25 °C and less than 0.1 ppm of $O_2$ and $H_2O$. The electrodes were separated (to prevent contact) by a Celgard polymer separator and submerged in an electrolyte solution of 1 M LiPF6 in 1:1:1 ratio of ethylene carbonate (EC): diethyl carbonate (DEC): dimethyl carbonate (DMC).

The working electrodes consisted of a multi-layered structure illustrated in the inset in Fig. 1. They were prepared by depositing a 15 nm of Ti as adhesion layer, 300nm of Cu as current collector, and an amorphous Si film on a (111) single crystal Si substrate of 400 μm thickness and 50.8 mm diameter. Approximately 300 nm of $SiO_2$ was thermally grown on all sides of the Si wafers prior to this deposition to create a barrier for Li diffusion and isolate the Si substrate from chemical reactions. The Ti and Cu films were deposited by e-beam evaporation technique under ultra-high vacuum (<~10-6 Torr) whereas the Si was deposited by RF-magnetron sputtering at 180 W power and less than 2 mTorr Ar pressure (using Lesker PVD Lab-18 from Kurt J. Lesker Inc., PA, USA). The sputtered Si films under these conditions are known to be amorphous [51], which is confirmed by an XRD analysis on the deposited films.

The electrochemical performance of the cell was controlled and monitored during electrochemical cycling by means of a Solarton 1470 E multistate. The state of stress in the film was measured in real time as the working electrode was lithiated and de-lithiated. The stress was determined by monitoring the curvature of the substrate using a multi-beam optical sensor (kSA-MOS from K-Space Associates, Inc., Dexter,

Michigan). The average in-plane film stress $\sigma$, was obtained from change in curvature $\kappa$ of the substrate as the film was lithiated using Stoney equation [52] [53],

$$\sigma = \sigma_r + \frac{E_s h_s^2 \kappa}{6 h_f (1-\nu_f)} \qquad (1)$$

where $E_s, \nu_s, h_s$ are the Young's modulus, Poisson's ratio, and thickness of substrate, respectively, $\sigma_r$ is the residual stress in the Si film due to sputter deposition. The residual stress was measured by tracking the substrate curvature before and after each film-deposition (Ti/Cu, Si) step, and values are listed in Table 1.

The film thickness $h_f$, varies during lithiation and delithiation. In calculating the stresses, the thickness was taken to vary according to relation $h_f = h_{f0}(1 + 2.7z)$, where $h_{f0}$ is the initial film thickness, and $z$ is the state of charge. The state of charge is computed from $z = c/c_{max}$, where $c$ is the ratio of the molar density of Li and Si $c = \rho_{Li}/\rho_{Si}$ with respect to the reference state, and $c_{max}$ is the concentration ratio in the fully lithiated state. The concentration $c$ is computed from the expression

$$c(t) = \frac{1}{\rho_{Si} F h_{f0}} \int_0^t I(t) dt \qquad (2)$$

where $I$ is the externally applied current density (current per unit electrode surface area), $\rho_{Si}$ is the molar density of the un-lithiated amorphous Si film and $F$ is the Faraday constant (the charge of 1 mol of electrons). reached at the fully lithiated state. This procedure assumes a 370 % volume expansion of the film at maximum capacity for lithiated *a-Si* [54] and assumes that the volume expansion is accommodated by increasing the film thickness, since lateral expansion is prevented by the substrate. It also neglects charge lost to the solid-electrolyte-interphase layer, and consequently provides an upper bound to the film thickness.

Two series of tests were conducted, which were designed to determine the mechanical and electrochemical response of the Si electrode. The first series of tests were intended primarily to determine the mechanical properties of the Si film. To this end, small working electrode thicknesses were used, and the films were lithiated at modest rates, with a view to minimizing concentration gradients through the film thickness. The films were subjected to cycles of lithiation and de-lithiation at constant externally applied current. The mechanical loading imposed during this test resembles cyclic thermal stresses induced by repeatedly heating and cooling the film: during lithiation; a state of compressive stress develops, deforming the film plastically; when the Li is subsequently removed, the stress becomes tensile. The test conditions were designed to ensure that the films remained intact and did not fracture on de-lithiation (the integrity of the films was verified subsequent to each test). The results of these tests then enable the elastic modulus and flow stress to be determined as functions of concentration. In addition, by varying the rate of lithiation, the strain rate sensitivity of plastic flow can be deduced. Finally, the voltage cycles measured in this experiment enable parameters controlling the free energy of mixing in the solution thermodynamics to be determined.

The relevant conditions for these mechanical tests are summarized in Table 1. In the first experiment a 127nm thick a-Si film (Cel151) was lithiated and delithiated at constant current densities of 5 µA/cm², 10 µA/cm², and 15 µA/cm² in the first, second, and remaining cycles, respectively; the current density and potential histories are shown in Fig. 3a and 3b respectively (blue curves). The cell was kept under open circuit condition for 5 minutes after every lithiation and delithiation step. In all cycles, lithiation was carried out until a lower cut-off voltage of 0.05 V vs. Li/Li+ to prevent crystallization of lithiated Si and delithiation was done until an upper cut-off voltage of 0.6 V vs. Li/Li+ to prevent film cracking. The corresponding stress evolution is shown in Fig. 3c.

In the second experiment a 103nm thick a-Si film (Cell 200) was lithiated and delithiated at a constant current density of 8.75 µA/cm$^2$, between 0.6 V vs. Li/Li+ and 0.05 V vs. Li/Li+, for ten cycles. A five minute open circuit condition was imposed after each lithiation and delithiation step. The current vs. time history is shown in Fig. 4 (blue curve) and the corresponding potential and stress histories are shown in Fig. 5 (blue). Note that the potential and stress are plotted against capacity (Li concentration); the area under the potential hysteresis loop in each cycle represents the total energy dissipated in lithiating and delithiating the film. Since capacity is also a measure of plastic strain in the Si film, the area of each stress hysteresis loop is proportional to the mechanical energy dissipation due to plastic deformation. These results will be discussed in more detail following the description of the model in Section 4.

A separate series of tests were conducted to determine the kinetic parameters (such as the diffusion coefficient) governing Li transport through the electrodes as well as the kinetics of the surface Li insertion reaction. Two experiments are typically used for this purpose, known as (i) Galvanostatic or (ii) Potentiostatic Intermittent Titration Technique (GITT/PITT). Both approaches start with an electrode at equilibrium, and then inject a small quantity of Li into the surface of the electrode at high rate. In GITT, a prescribed current is used for the Li insertion; while in PITT a step increment in potential is applied. Both tests induce a non-uniform concentration distribution in the film, and the transport properties are determined by monitoring the rate at which the system subsequently returns to equilibrium. The kinetic parameters of interest can then be extracted from these tests with the aid of a theoretical model.

The standard approach to modeling interrupted titration experiments neglects the influence of stress on Li transport and so some modifications are necessary in order to apply the procedure to the electrodes tested here. We selected PITT experiments over GITT with a view to minimizing the effects of stress. The films were first lithiated to a pre-determined voltage that was sufficient to cause the electrode material to yield in compression. Additional increments in potential were then applied in a direction chosen to increase the compressive stress. Since the flow stress varies only slowly with lithiation rate and Li concentration, the stresses remain approximately constant during the voltage step. After a succession of tests of this type, the process was repeated with tensile loading.

PITT experiments were conducted on a 104 nm a-Si thin-film electrode in a beaker cell against a lithium metal reference and counter electrode. Lithium insertion was carried out galvanostatically at 2.5 µA/cm$^2$ (geometric area) until the electrode potential reached 0.4 V *vs.* Li/Li$^+$. This was followed by potentiostatic steps at 0.4 V, 0.35 V, 0.3 V, … (*i.e.*, 50 mV increments till 0.05 V *vs.* Li/Li$^+$). Each potentiostatic step was carried out until the current decayed to less than 0.25 µA/cm$^2$. These were followed by identical potentiostatic steps in the delithiation direction (*i.e.*, potential steps at 50 mV increments till 1.2 V *vs.* Li/Li$^+$). The history of the potential and current density during a PITT experiment are shown in Fig. 6a and 6b respectively (blue curves). The stress in the Si thin-film electrode was continuously measured *via* the substrate-curvature-monitoring technique, which is shown in Fig. 6c. It should be noted that the PITT experiments described herein were carried out on an as-prepared Si thin-film electrode.

Although this PITT procedure minimizes the effects of stress, it does not eliminate them completely. Rather than using the standard models of PITT experiments to extract material parameters, therefore, a full nonlinear simulation of the experiment was conducted, and a value for diffusivity was also determined that provides the best fit to experimental measurements. The approach to modeling the experiments is described in more detail in the next section.

## 3. Model of a half-cell with a thin-film Si anode

Fig. 2 shows an idealized model of the half-cell that was characterized experimentally. It consists of a thin Si film, with initial thickness (in the un-lithiated state) $H$, which is bonded to a rigid substrate; together with a Li counter-electrode. The two electrodes are separated by an electrolyte, and are connected through an external circuit, which maintains either a prescribed history current flow $I$ or a prescribed voltage drop $V$ across the electrodes. Note that in a half-cell with a Li counter-electrode, Si is the positive electrode, in contrast to a full cell, in which Si is the negative electrode.

When the cell is first assembled, the Si electrode is free of Li and has a positive electric potential with respect to the Li counter-electrode. Before lithiation, the electrode is subjected to a biaxial residual stress $\sigma_r$. Connecting the external circuit between the electrodes permits current to flow from the Si electrode to the counter-electrode, which enables the electrochemical reactions to take place at the interfaces between the electrolyte and the electrode. At the surface of the Li electrode, the reaction $Li \rightarrow Li^+ + e^-$ removes Li atoms from the electrode surface, which enter the electrolyte as Li ions. At the surface of the Si electrode, the reaction $Li^+ + e^- \rightarrow Li$ takes place; Li atoms are inserted into the electrode, and subsequently diffuse through the thickness of the film. In addition, a variety of side-reactions may occur, in which the electrolyte is reduced to form a passivating solid electrolyte interphase (SEI) layer on the electrode surface. In these reactions, additional electrons combine with Li ions in the electrolyte, but the Li involved in these reactions cannot be recovered. The rate of these side reactions progressively decreases as the passivating layer increases in thickness. In subsequent discussions, we will denote the electric current density per unit area of Si electrode surface resulting from the Li insertion reaction by $I_R$, and the additional current density associated with SEI formation and other side reactions by $I_S$. We adopt a sign convention wherein positive current flow denotes a flow of electric current through the external circuit from the Si electrode to the Li counter-electrode.

As Li is inserted into the Si electrode, its volume increases. The substrate prevents the electrode from expanding laterally, so a large biaxial compressive stress develops in the film during the first lithiation cycle. The stress is sufficient to cause permanent plastic deformation in the film, and the electrode accommodates the Li insertion by increasing its thickness. At full capacity, the film thickness increases by nearly a factor of 3, so it is important to account for finite geometry changes when modeling Li transport, stress, and deformation in the electrode.

As the Li concentration in the electrode increases, its electric potential relative to the counter-electrode decreases. When it drops to a pre-determined value, the external current is reversed, and Li is removed from the electrode. This induces tensile stress, which reverses the plastic strain. As Li is removed from the Si electrode its potential relative to the counter-electrode increases, and eventually reaches the initial value. In our simplest tests, the electrode is subjected to a number of successive cycles of this kind with a prescribed total current density. In more complex tests, such as the PITT experiments, a prescribed history of voltage $V$ is applied, and the current is measured.

The goal of our model is to predict the mechanical displacement and stress fields in the electrode; the variation of Li concentration through the film thickness; the electric potential difference $V$ between the electrodes, and the total current flow $I = I_R + I_S$ through the external circuit per unit surface area of electrode.

For this purpose, we adapt a continuum model of an electrochemical Li ion cell previously described in detail in [25]. We simplify calculations by neglecting electrical resistance (electrons flow freely); assuming infinitely fast Li ion transport through the electrolyte; assuming that the electrolyte remains electrically neutral, and neglecting the potential drop across the counter-electrode/electrolyte interface. The entire potential difference $V$ between the two electrodes thus appears across the Si/electrolyte interface. We then focus attention only on the transport, chemistry and deformation in the thin film Si electrode, the electrochemical reaction at the Si/electrolyte interface that is responsible for inserting Li into the electrode; and the side reaction associated with SEI formation.

We idealize the electrode material as an amorphous network of atoms, with molar density $\rho_{Si}$ prior to lithiation. We take the stress free Si as the reference configuration. When the electrode is lithiated, a molar density $\rho$ of Li atoms per unit reference volume is accommodated within interstitial sites of the reference lattice. In the following calculations, it is more convenient to characterize the Li content by the molar Li concentration $c = \rho_{Li} / \rho_{Si}$. We next proceed to describe separately the equations governing mechanical deformation in the electrode; Li thermodynamics and transport through the film; and electrochemical reactions at the Li/electrolyte interface.

## 3.1 Mechanical Deformation

As Li is inserted into the Si electrode, it increases its thickness. As a result, a material particle that starts at position $X$ above the substrate will move to a new position $x$, with displacement $u = x - X$. The deformation of the film can be characterized by the logarithmic 'true' strains

$$\epsilon_{xx} = \log\left(1 + \frac{\partial u}{\partial X}\right) \qquad \epsilon_{yy} = \epsilon_{zz} = 0 \tag{3}$$

The strain in a representative volume element occurs by three processes, which can be visualized as occurring in sequence. First, *Li* insertion causes the stress-free *Si* network to increase its volume. We assume that the volume change is proportional to the Li molar concentration, so $dV/dV_0 = 1 + \beta c$, with a corresponding isotropic true strain $\log(1 + \beta c)/3$. The expansion is accompanied by a small elastic distortion of the Si network (which induces stress), which we characterize by an infinitesimal elastic strain $\varepsilon_{xx}^e, \varepsilon_{yy}^e = \varepsilon_{zz}^e$. In addition, the solid experiences an irreversible, volume preserving plastic deformation, which we characterize by a plastic stretch ratio $\lambda^p$ in the plane of the film. Since the substrate prevents the film from expanding in its own plane, the strains are related by the compatibility conditions

$$\frac{1}{3}\log(1+\beta c) - 2\log(\lambda^p) + \epsilon_{xx}^e = \log\left(1 + \frac{\partial u}{\partial X}\right) \tag{4}$$

$$\frac{1}{3}\log(1+\beta c) + \log(\lambda^p) + \epsilon_{yy}^e = \frac{1}{3}\log(1+\beta c) + \log(\lambda^p) + \epsilon_{zz}^e = 0 \tag{5}$$

We characterize the stress state in the film by the Cauchy stress $\sigma_{yy} = \sigma_{zz} = \sigma(x)$. We take the stress to be related to the elastic strains by a concentration dependent biaxial modulus, so that

$$\sigma = M(c)\epsilon_{yy}^e \tag{6}$$

We relate the modulus to the concentration by

$$M = M_0 + M_1 \log\left(1 + \frac{c}{c_0}\right) \tag{7}$$

where $M_0, M_1$ are material constants and $c_0$ is the initial Li concentration. Note that the elastic constitutive law used here differs from [25], which prescribed a linear relationship between Kirchhoff stress and elastic strain. A logarithmic constitutive law for the biaxial modulus has been chosen, taking into account the measurements published in [22].

The plastic deformation is characterized by a viscoplastic constitutive equation relating the plastic stretch rate to stress

$$\frac{\dot{\lambda}^p}{\lambda^p} = \begin{cases} 0 & |\sigma| \leq \sigma_0 \\ \frac{\dot{\varepsilon}_0}{2}\left(\frac{|\sigma|}{\sigma_0}-1\right)^m \frac{\sigma}{|\sigma|} & |\sigma| \geq \sigma_0 \end{cases} \quad (8)$$

where $\sigma_0$ is a concentration dependent yield stress; $\dot{\varepsilon}_0$ is a characteristic strain rate, and $m$ a stress exponent. The latter two parameters are taken to be constant. The yield stress constitutive law is assumed to be a linear function of lithium concentration

$$\sigma_0(c) = s_0 + s_1(c - c_0) \quad (9)$$

where $s_0 = s|_{c=c_0}$ and $s_1$ are constant material parameters.

## 3.2 Solution thermodynamics and Li transport

We adapt the model outlined in [25] to describe Li transport and alloying in the Si electrode. Diffusion of Li through the Si electrode is driven by a chemical potential

$$\mu = \mu^\Theta + RT \ln \frac{\gamma c}{c_{max} - c} - \frac{\sigma^2}{\rho_{Si}} \frac{\partial}{\partial c}\left(\frac{1}{M}\right) - \frac{2\sigma\beta}{3(1+\beta c)\rho_{Si}} \quad (10)$$

where $R$ is the gas constant and $T$ is temperature. In eq. (10) the first term is a reference chemical potential; while the second term quantifies the chemical interaction of Li with the Si host matrix. With $\gamma = 1$, this term accounts for entropic contributions from the Li concentration and the host sites ($c_{max}$ is the maximum Li molar concentration when all host sites have been filled). The concentration dependent activity coefficient $\gamma$ quantifies the free energy change resulting from bonding between Li and the host. Guided by Verbrugge and Koch [55], we use the following form for the activity coefficient

$$RT \ln \gamma = \sum_{n=2}^{N} \Omega_n n \left(\frac{c}{c_{max}}\right)^{n-1} \quad (11)$$

where $\Omega_n$ are a set of coefficients that must be determined experimentally. Finally the last two terms in eq. (10) are the contributions to the chemical potential from stress, which have been derived by several authors [56] [57]. The variation of chemical potential with Li concentration used here differs from ref. [25], which included only the entropic contribution to the free energy. The additional terms used here are necessary to characterize the variation of open circuit potential with concentration correctly, and have a significant influence on transport under non-equilibrium conditions.

The molar flux of Li crossing a surface with unit area in the plane of the film is related to the chemical potential gradient through Fick's law

$$j_X = -\frac{D}{RT}\rho_{Si} c \frac{\partial \mu}{\partial x} \quad (12)$$

where $D$ is the diffusion coefficient. Note that the spatial derivative is used in eq. (12), which differs from ref [25]. The spatial description ensures that the effective diffusion coefficient is independent of deformation. Mass conservation requires that

$$\rho_{Si} \frac{dc}{dt} = -\frac{dj_X}{dX} \quad (13)$$

Finally, the Li flux at the upper and lower surfaces of the film satisfies
$$j_X = 0 \quad (X = 0) \qquad j_X = I_R / F \quad (X = H)$$

where $F$ is the Faraday constant (the absolute value of the charge of one mol of electrons), and $I_R$ is the electric current associated with the charge transfer reaction that accompanies Li insertion into the electrode. Separate constitutive equations, to be listed in the next section, govern the relation between $I_R$ and the electric potential difference across the Si electrode/electrolyte interface.

### 3.3 Surface electrochemical reactions

Two separate electrochemical reactions are assumed to take place at the Si/electrolyte interface. In the first, Li ions in the electrolyte combine with electrons and are inserted into host sites $S$ in the Si network, with the reaction

$$Li^+ + e^- + S \rightarrow \left[ Li^{+\delta} - S^{-\delta} \right] \quad (14)$$

Here, $S$ denotes available host sites. We quantify the rate of this reaction by the phenomenological Butler-Volmer equation, which relates the current density $I_R$ to the over-potential at the solid/electrolyte interface as follows

$$I_R = i_0 \left[ \exp\left( \frac{\alpha F \eta}{RT} \right) - \exp\left( -\frac{(1-\alpha) F \eta}{RT} \right) \right] \quad (15)$$

Here, $0 < \alpha < 1$ is a phenomenological constant, $\eta = V - U_0$ is the 'overpotential' applied across the interface – the difference between the externally applied potential $V$ and the open circuit potential $U_0$. A straightforward thermodynamic argument shows [58] that the open-circuit potential must be related to the chemical potential of Li at the surface of the electrode by

$$U_0 = -\frac{\mu}{F} = -\frac{\mu^\Theta}{F} - \frac{RT}{F} \ln \frac{\gamma c}{c_{max} - c} + \frac{\sigma^2}{F \rho_{Si}} \frac{\partial}{\partial c}\left( \frac{1}{M} \right) + \frac{2\beta\sigma}{3F(1+\beta c)\rho_{Si}} \quad (16)$$

The factor $i_0$ in eq. (15) is the exchange current density (which, indirectly, characterizes the apparent resistance of the interface to electric current flow). The exchange current density depends on the Li concentration at the surface of the electrode, and the Li concentration $\rho_{Li+}$ of Li$^+$ ions in the electrolyte adjacent to the interface through

$$i_0 = F \left[ k_c \rho_{Li+} \left( 1 - \frac{c}{c_{max}} \right) \right]^\alpha \left( k_a \frac{c}{c_{max}} \right)^{(1-\alpha)} \quad (17)$$

Here, $k_c$ and $k_a$ represent concentration dependent cathodic and anodic rate constants. The constants cannot be determined uniquely from our experimental data, so we use the following expression to fit the variation of a combination of reaction rates with concentration

$$\left( \rho_{Li+} k_c \right)^\alpha k_a^{(1-\alpha)} = k_0 + k_1 \sin\left( \frac{\pi}{2} \frac{c}{c_{max}} \right) \quad (18)$$

where $k_0$ and $k_1$ are constants that must be fit to experimental data.

In addition to the reaction that leads to Li insertion into the electrode, additional reactions occur at the negative electrode/electrolyte interface whenever the electrode potential falls below the value necessary to reduce the electrolyte. The product of the electrolyte decomposition forms a solid layer on the surface of the active material. This SEI layer is permeable to lithium ions, but has a high resistivity, and electrolyte can diffuse only slowly through the SEI. The rate of the side reactions thus decreases as the SEI layer increases in thickness. This SEI formation represents a significant source of capacity loss in the thin film electrodes tested in our study, and must be modeled explicitly. We attempted to fit our experimental data with standard models of SEI formation [46] [47] [48]. These generally predict that the SEI thickness and total capacity loss vary with time as $t^{1/2}$ after an initial transient. We found that these relations do not fit our experimental data. As previously reported by Nadimpalli *et al* [51], we observe an initial rapid loss of capacity during the first lithiation cycle, but the rate of capacity loss during a small number of subsequent cycles is difficult to detect. Standard SEI models give a good fit to capacity loss measurements over a long time period (e.g. [48]) but underestimate the large first cycle capacity loss. We have therefore used a simple phenomenological relation to fit our data. To account for side reactions, the total current through the external circuit is divided into two contributions

$$I = I_R + I_S \tag{19}$$

where $I$ is the total externally applied current, $I_R$ is the current supplied to the desired Faradaic reaction calculated using eq. (15), and $I_S$ is the current going into SEI formation. The electrolyte reduction is assumed to occur at the surface of the electrode, and to follow Tafel kinetics, with an equilibrium potential $U_{0,SEI}$ of 0.8 V vs. Li/Li+ [12]. The current density associated with the SEI reaction is expressed as

$$I_S = -i_{0,SEI}\left(1 - \frac{Q_{loss}}{Q_{SEI}}\right)\exp\left(-\frac{2\alpha_{side}F}{RT}(V - U_{0SEI})\right) \tag{20}$$

where $i_{0,SEI}$ and $Q_{SEI}$ are phenomenological constants, $\alpha_{side}$ is the apparent transfer coefficient and $V$ is the electrode potential, and

$$Q_{loss} = -\int_0^t I_S dt \tag{21}$$

is the total charge per unit area of surface lost to the side reaction.

### 3.4. Numerical solution of governing equations

The equations listed in the preceding section are in general too complex to be solved analytically. There is no difficulty in obtaining numerical solutions, however. We have followed two approaches to solving the equations here. For the limiting case where the film is lithiated very slowly compared to the characteristic time associated with Li transport through the thickness of the film, the analysis can be simplified by assuming that the Li concentration remains uniform through the electrode thickness. In this case, the Li concentration is related to the total current by

$$\frac{\partial c}{\partial t} = -\frac{I_R}{F \rho_{Si} H} \tag{22}$$

where $H$ is the initial film thickness, and $I_R$ is related to the total external current and potential by eqs (15), (19), and (20). The stress in the film is then uniform and can be computed by integrating the ODE

$$\frac{\partial}{\partial t}\left(\frac{\sigma}{M}\right) = -\frac{\beta}{3(1+\beta c)}\frac{\partial c}{\partial t} - \frac{\dot{\epsilon}_0}{2}\left(\frac{|\sigma|}{\sigma_0} - 1\right)^m \frac{\sigma}{|\sigma|} \tag{23}$$

If significant concentration gradients develop through the film thickness, a more sophisticated procedure must be used to determine the variations of stress and concentration through the film thickness. We have used a simple one-dimensional finite element method for this purpose, which is summarized briefly in Appendix A.

## 4. Determining values for material parameters

The model of the electrochemical-mechanical problem involves a large number of parameters. Where possible, parameters were determined from literature data. In particular, the molar density and elastic properties of un-lithiated amorphous Si were determined from refs [53], [59], [60]; while the volumetric expansion of Si with Li concentration was taken from ref [61]. Values for these properties are listed in Table 2. In the following sections, we describe briefly the procedure that was used to fit parameters to experimental measurements.

### 4.1 Elastic and plastic properties of a-Si

The material parameters governing the mechanical response of the Si electrode were fit to experimental measurements in which a 100nm thick electrode was lithiated and de-lithiated at a constant current density, as described in Section 3. The conditions of the experiment are summarized in Table 1. The electrode thickness in this experiment is sufficiently small to ensure that the concentration remains uniform. The material is therefore effectively subjected to a prescribed cycle of externally applied strain as Li is inserted and removed from the film. By measuring the resulting cycle of stress, it is possible to determine the parameters $(M_0, M_1)$ characterizing the variation of elastic modulus with concentration in eq. (7), together with the parameters $(\dot{\varepsilon}_0, s_0, s_1, m)$ that govern the rate dependent plastic flow in the film, as defined in eqs. (8) and (9).

These parameters were fit to the cycles of stress shown in Figure 3c (red curve), which were obtained from the measurements corresponding to Cell151 in Table 1. The elastic constants $(M_0, M_1)$ were fit to the slope of the stress-v-capacity curves at the point where the direction of the external current was reversed. The parameters $(\dot{\varepsilon}_0, s_0, s_1)$ were fit to obtain the correct variation of flow stress with capacity. The values that best fit the experiment are listed in Table 2. Since the flow stress did not change significantly as the lithiation rate was increased in the successive cycles of this test, it appears that the Si flow stress is only weakly strain rate sensitive. Any value for the stress exponent $m>50$ will fit the experimental data. We have used $m=50$ in all the remaining computations reported in this paper.

### 4.2 Parameters characterizing solution thermodynamics and Li insertion reactions

The electrochemical performance of the working electrode is determined primarily by the concentration dependent activation coefficient $\gamma$ in equations (10) and (11), together with the variation of exchange current density defined in eqs. (15), (17) and (18). These are parameterized by the coefficients $\Omega_n$ in eq. (11) (which quantify the free energy of mixing), and the coefficients $(k_0, k_1)$ in eq. (18) (which characterize the variation of exchange current density with concentration).

In principle, the coefficients $\Omega_n$ can be determined through equation (16) by measuring simultaneously the variation of rest potential and stress with concentration. The rest potential is difficult to determine accurately, however, because the potential vs. current curves in experiments differ during lithiation and de-lithiation. We have therefore used literature data (from GITT experiments [41] and from measurements and first principle calculations [33]) to determine $\Omega_n$ and $\Delta\phi_0^\Theta$. The resulting fit is shown in Fig 7, which displays the predicted open-circuit potential for a stress-free electrode as a function of concentration. The experimental data from ref [41] and the *ab-initio* predictions from ref [33] are shown for comparison.

The coefficients $(k_0, k_1)$ in eq. (18) were determined from our PITT experiments. The values were selected to fit the peak values of the current at the beginning of each voltage step. The resulting values, listed in Table 2, allowed an accurate prediction of the quick drop or increase of the potential at the beginning of a charge or discharge cycle in galvanostatic experiments (as shown in Fig. 3b and 5a for Cell 151 and Cell 200 respectively).

Finally, the exchange current density $i_{0SEI}$ in our model of capacity loss to SEI formation (eq. (20)) was determined by fitting the predicted time variation of voltage for cell 151 for the four successive charge/discharge cycles to experiment. The total capacity $Q_{SEI}$ loss was taken from experimental measurements previously reported in ref [51].

### 4.3 Lithium diffusivity

The potentiostatic intermittent titration technique (PITT) is an electroanalytical method developed by Weppner *et al* [49] [50] to study diffusion of the mobile species (such as lithium) across the thinnest dimension of a solid electrode in a slab geometry. In the standard PITT measurement, a step change in voltage $\delta V$ is applied to the electrode, and the subsequent transient current $I(t)$ is measured. If transport in the electrode obeys Fick's law

$$\frac{\partial c}{\partial t} = D\frac{\partial^2 c}{\partial x^2} \tag{24}$$

and the Li insertion reaction occurs sufficiently rapidly to ensure that transport in the electrode is the rate limiting process, then the transient current in an electrode with thickness $h$ can be approximated by the relation

$$I(t) = \frac{2QD}{h^2}\exp\left(-\frac{\pi^2 Dt}{4h^2}\right) \tag{25}$$

where

$$Q = \int_0^\infty I(t)dt \tag{26}$$

The diffusion coefficient can be extracted by fitting the predictions of eq. (25) to experiment. In the system tested here, however, the transient current does not follow the behavior predicted by eq. (25). This is partly because the Li insertion reaction kinetics cannot be neglected (Verbrugge *et al* [39] provide a correction to (25) to account for this); and partly because the chemical potential in eq. (10) departs from ideal behavior. For small perturbations in concentration $\delta c$ relative to an initial concentration $c$, eqs. (10), (12) and (13) can be combined and linearized to yield

$$\frac{\partial \delta c}{\partial t} = \tilde{D}\frac{\partial^2 \delta c}{\partial x^2} \tag{27}$$

where

$$\tilde{D} = D\frac{c}{RT}\left[\left.\frac{\partial \mu}{\partial c}\right|_\sigma + \left.\frac{\partial \mu}{\partial \sigma}\right|_c \frac{\partial \sigma}{\partial \delta c}\right] = \frac{D}{RT}\left[\left(\frac{c_{max}}{c_{max} - c} + \frac{\partial \ln \gamma}{\partial \ln c}\right) + c\left.\frac{\partial \mu}{\partial \sigma}\right|_c \frac{\partial \sigma}{\partial \delta c}\right] \tag{28}$$

is an apparent diffusion coefficient. The apparent diffusion coefficient can be extracted from a PITT measurement, and the intrinsic diffusivity can be determined if the enhancement factor $\tilde{D}/D$ can be found. The first two terms in the correction to the diffusion coefficient arise from the non-ideal free energy of mixing. The first term provides a significant contribution when the Li concentration is close to the saturation value $c_{max}$. In addition, because the chemical potential $\mu$ depends on stress, the enhancement factor is history dependent. If the increment in concentration $\delta c$ is such that the electrode deforms elastically, then $\partial \sigma / \partial \delta c \approx M\beta$, whereas for plastic loading $\partial \sigma / \partial \delta c \approx s_1$. The effects of stress can thus be minimized by ensuring that the voltage step $\delta V$ tends to deform the electrode plastically.

We have accordingly used two approaches to estimate values for the diffusion coefficient from our experiments. In the first approach, values for $\tilde{D}$ were estimated by fitting eq. (25) to the measured transient current curves, and $D$ was determined by correcting the measurements using eq. (28). The results of these calculations are displayed as symbols in Fig 8. In this approach, each successive experiment yields a separate value for $D$. In the second approach, a value for $D$ was determined by fitting full numerical simulations of the PITT experiment to the experimental data. This approach accounts rigorously for departures from the ideal diffusion model resulting from stress and the free energy of mixing, and also accounts for nonlinearities arising from the Li insertion kinetics. A single diffusion coefficient of $D = 10^{-19}$ m$^2$s$^{-1}$ was found to predict the experimentally measured transients in the PITT experiments over the full range of Li concentration. This value also corresponds approximately to the average value of the diffusion coefficients determined from the classical model of the PITT experiment.

## 5. Discussion

We proceed to compare the predictions of our model with experimental measurements. Fig. 3 shows the predictions and experimental data for Cell 151. The data from this test were used to determine material parameters. Fig 3a shows the variation of externally applied current $I$ with time for this experiment: the electrode was lithiated and de-lithiated in five successive cycles with a progressively increasing applied current, which effectively exposes the electrode material to an externally applied cycle of volumetric strain at a progressively increasing rate. The figure also shows the predicted variation of the lithiation current $I_R$ with time in this experiment: the difference $I_S = I - I_R$ is expended in forming SEI on the electrode surface. The total charge loss of 0.014 mAh/cm$^2$ is comparable to that measured by Nadimpalli *et al* [51].

The measured and predicted variations of stress in Cell 151 are compared in Fig 3c. Note that both experimental measurements and computations show the apparent true stress (the total film force, obtained by integrating the computed Cauchy stress through the current film thickness is divided by the nominal current electrode thickness computed according to eq(2)), and also include the initial residual stress. During cycles 2-5, the measured and predicted stresses are in good qualitative agreement: the predictions capture the variation of modulus and flow stress with Li concentration and lithiation rate. The detailed features on the experimental stress curves that are not captured perfectly by the model, but these features are not repeatable (see data for cell 200 shown in Fig 5b, for example). The most significant discrepancy between predicted and measured behavior occurs during the first lithiation cycle. The model predicts elastic behavior up to the initial yield (the stress-v-capacity curve is somewhat lower than the elastic modulus because of charge lost to the SEI) but the slope of the stress-v-capacity curve measured experimentally is consistently well below model predictions. It is possible that some structural rearrangement, akin to a phase transition in crystalline Si, occurs during the initial lithiation of a-Si. After the initial lithiation cycle, however, model predictions are in good agreement with experiment.

Our experimental data suggests that elastic biaxial modulus varies from 102.6 GPa in the un-lithiated electrode to $64.1 \pm 19.2$ $GPa$ at full charge capacity, which is in good agreement with values previously reported in [31]. The flow stress measured here is comparable to values measured in [25]. The value $m=50$ found for the stress exponent is substantially greater than values reported previously, however. For example, ref [25] used a value $m=4$ to fit their data, but this value was fit to only a single charge-discharge cycle at a fixed current density, and so did not provide an accurate measure of the stress exponent. Soni et [62] found $m=5$ fit stress measurements in experiments similar to those described here, but in their experiments the films were found to fracture, which would tend to enhance the variation of the average stress with charging rate.

Recent atomistic simulations of Li insertion and plasticity in amorphous Si [32] suggest that a more sophisticated model of the kinematics of Li insertion and plasticity may be required than the multiplicative decomposition of deformation gradients that leads to eqs. (4) and (5). Our constitutive equations assume that volumetric strains in the amorphous Si network depend only on the Li concentration, and deviatoric (volume preserving) plastic strains occur only as a result of deviatoric (shear) stresses. Motivated by atomistic simulations, Zhao *et al* [29] have proposed a more sophisticated constitutive model in which plasticity and Li insertion are fully coupled. The experimental measurements reported here cannot determine whether this more sophisticated treatment is necessary to predict accurately the variation of stress in our thin-film electrodes, since both approaches will fit our experimental data with appropriate material parameters. Experimental measurements in which the stress in the a-Si can be varied at fixed Li concentration, for example using micro-pillar indentation tests, would be necessary to resolve this question.

The measured and predicted voltage vs. time curves for Cell 151 are compared in Fig. 3b. The predicted voltage differs slightly from experiment during the initial cycle of lithiation (the discrepancy may be related to the difference between measured and predicted stress, since the cell potential is stress dependent). After the initial cycle, however, the model predicts voltages that are in excellent agreement with experiment. It should be noted that although (16) contains a number of adjustable parameters through the concentration dependence of the activation coefficient (defined in eq. (11)), these parameters were fit to ab-initio simulations reported in ref. [33].

We have validated the model by predicting the stress and voltage for a second cell, using the same material parameters determined from cell 151. Cell 200, which had an electrode with thickness 123nm, was subjected to 10 successive charge/discharge cycles at a constant current density (test conditions are listed in Table 1). The variation of current density with time (and the predicted current $I_R$ associated with Li insertion) is shown in Fig. 4, and the measured and predicted mechanical and electrical responses of this electrode are compared in Fig 5. The results are displayed as the variation of stress or potential with capacity, to enable behavior under successive cycles to be compared. As before, there is a discrepancy between measured and predicted stress during the initial lithiation. For subsequent cycles, model predictions are in good agreement with experiment, and capture the evolution of stress with successive cycles of charge. Similarly, the predicted voltage-v-capacity curve exceeds the experimentally measured values for the first three cycles, but is in good agreement for cycles 4-10.

Finally, we compare the measured and predicted behavior during PITT experiments in Fig. 6, in which the electrode was subjected to a series of step changes in voltage. Fig 9 and 10 show the predicted variation of Li concentration, stress, and plastic strain through the thickness of the film following a representative lithiation step (Fig 9) and a de-lithiation step (Fig 10). A step reduction in voltage causes a transient increase in Li concentration at the surface of the film, which subsequently propagates through the film thickness. The rate of decay of this transient is controlled by the diffusion coefficient, The predicted variations of stress and plastic strain through the film thickness are of particular interest, since stress gradients give rise to additional driving forces for diffusion and cause the transient current measured in the PITT experiment to deviate from the predictions of the idealized model in eq. (25) . The experiments were designed to minimize the effects of stress by selecting the voltage steps so as to deform the film plastically. Model predictions suggest that an approximately 10% variation in stress occurs through the thickness of the film during each transient, so the term involving stress in eq. (28) can be neglected.

The predicted average through-thickness stress and the magnitude of the current following each voltage step are compared to experimental measurements in Fig 6b, while Fig 11 shows a more detailed comparison of the measured and predicted transient current following selected voltage steps during both lithiation and de-lithiation. The model predicts accurately the step increase in current following each voltage step and for most voltage steps also predicts correctly the subsequent transient decay. There are some discrepancies between measured and predicted relaxation curves following voltage steps from 0.2 V to 0.05 V during the lithiation phase of the cycle, but the transient currents measured experimentally for these voltage steps differ substantially from the expected exponential decay, suggesting that some physical process not considered in our model may begin to play a role at low voltages.

The predicted variation of stress with time is in qualitative agreement with the PITT experiments. The model correctly estimates the magnitude of the short-term transient stress variations following a voltage jump.

## 6. Conclusions

A combination of experimental measurements and numerical simulations were used to characterize the mechanical and electrochemical response of thin-film amorphous Si electrodes during cyclic lithiation. In the experiments, films were subjected to repeated cycles of lithiation and de-lithiation at either prescribed current density, or with prescribed increments of applied external potential. The variation of stress and current (or potential) were measured in-situ. The experiments were idealized by extending the continuum model from ref [25] to include a more accurate description of Li solution thermodynamics and transport, as well as a simple treatment of the capacity loss resulting from formation of the solid electrolyte interphase on the electrode surface.

The comparison of theory and experiment enabled material parameters characterizing both the mechanical and electrochemical response of amorphous Si to be extracted from the experiments; resulting values are listed in Table 2. The variation of elastic biaxial modulus was fit using a logarithmic variation with Li concentration, and was found to vary from 102.6 GPa in the un-lithiated electrode to $64.1 \pm 19.2$ $GPa$ at full charge capacity (the Young modulus varies varies from 80 GPa to $50 \pm 15$ $GPa$). These values are in good agreement with previously reported data [31]. The inelastic response was approximated using a power-law rate dependent viscoplastic constitutive equation, with a linear concentration dependent flow stress varying from $0.49 \pm 0.08$ $GPa$ at zero concentration to $0.23 \pm 0.08$ $GPa$ at full capacity, and a stress exponent of 50.

The Li solution thermodynamics were characterized in our model by a series expansion of the variation of open circuit potential for a stress free electrode as a function of concentration. The variation of potential with concentration predicted by *ab-initio* computations reported in [33] were found to give a good match to experimentally measured potential-v-capacity curves. PITT measurements, in combination with numerical simulations, were also used to determine the diffusion coefficient for Li in a-Si. Experiments were best fit with a diffusion coefficient $10^{-19}$ m$^2$s$^{-1}$, which is comparable to, but on the low end of the range $10^{-16}$ -$10^{-10} cm^2 s^{-1}$ measured in previous experiments [37] [38] [39] [40].

The experiments were also used to estimate the variation of exchange-current density in the Li insertion reaction with concentration. The exchange current density was found to be $i_0 \sim O(10 \text{ A/cm}^2)$, comparable to the one obtained by Chandrasekaran et al. [44], while the values reported in literature for lithium–silicon battery systems [63] [64] [65] vary by several orders of magnitude.

## 7. Acknowledgments

This work was supported by the U.S. Department of Energy through DOE EPSCoR Implementation Grant no. DE-SC0007074.

## Appendix A: Numerical procedure

In this appendix we outline briefly the one-dimensional finite element method that was used to compute the variation of concentration and stress in the electrode. Following the standard procedure, we interpolate the variation of concentration $c$ and chemical potential $\mu$ between a set of discrete values $c^a, \mu^a$ defined at a set of discrete points $X_a$ in the reference configuration. The values of $c, \mu$ over a representative interval $X_{a-1} < X < X_a$ are linearly interpolated as

$$c = N^{a-1}(X)c^{a-1} + N^a(X)c^a$$
$$\mu = N^{a-1}(X)\mu^{a-1} + N^a(X)\mu^a \tag{29}$$

through the shape functions

$$N^{a-1}(X) = \frac{X - X_a}{X_{a-1} - X_a} \qquad N^a(X) = \frac{X - X_{a-1}}{X_a - X_{a-1}} \tag{30}$$

The discrete values of $c^a, \mu^a$ are known at the time $t_n$, and must be computed at time $t_{n+1} = t_n + \Delta t$. To this end the diffusion equations are written in an equivalent integral form suitable for finite element discretization. For small elastic strains, the diffusion equations, together with the boundary conditions can be expressed in weak form as

$$\int_0^H \frac{\partial c}{\partial t} \delta c \, dX - \int_0^H \frac{D}{\beta RT} \frac{\partial \mu}{\partial X} \frac{\partial \delta c}{\partial X} dX = j_X(\text{H})\delta c(\text{H}) \tag{31}$$

where $j_X(\text{H})$ is the flux imposed on the top surface of the film. Similarly, the governing equation for the chemical potential is written as

$$\int_0^H \left[ \mu \delta \mu - \left( \mu^\Theta + RT \ln \frac{\gamma c}{c_{max} - c} - \frac{\sigma^2}{\rho_{Si}} \frac{\partial}{\partial c}\left(\frac{1}{M}\right) - \frac{2\sigma\beta}{3(1+\beta c)\rho_{Si}} \right) \delta \mu \right] dX = 0 \tag{32}$$

The variations of concentration and chemical potential $\delta c = N^b \delta c^b$, $\delta \mu = N^b \delta \mu^b$ are interpolated using the equations (29), and a forward-Euler scheme is used to integrate (31) with respect to time. Substituting the interpolations into (31) and (32) and noting that the equations must be satisfied for all $\delta c^b$, $\delta \mu^b$ yielding to a discrete system of nonlinear equations

$$R_a^c = \int_0^H \left[ \frac{\Delta c}{\Delta t} N^a - \frac{D}{\beta RT} \frac{\partial \mu}{\partial X} \frac{\partial N^a}{\partial X} \right] dX - j_X(\text{H})N^a(\text{H}) = 0 \tag{33}$$

$$R_a^\mu = \int_0^H \mu N^a dX - \int_0^H \left[ \mu^\Theta + RT \ln \frac{\gamma c}{c_{max} - c} - \frac{\sigma^2}{\rho_{Si}} \frac{\partial}{\partial c}\left(\frac{1}{M}\right) - \frac{2\sigma\beta}{3(1+\beta c)\rho_{Si}} \right] N^a dX = 0 \quad (34)$$

where $\Delta c = N^a \Delta c^a$ denotes the change in concentration during the time interval $\Delta t = t^{n+1} - t^n$, while $\mu = N^a \mu^a$ and $c = N^a c^a$ are the chemical potential and the concentration at time $t^{n+1}$.

The system of equations (33)-(34) may be solved by Newton-Raphson iteration for the unknown values $(\mu^a, c^a)$, $a = 1,\ldots k$, where $k$ is the number of nodes of the finite elements discretization. At the generic iteration the current approximation of $(\mu^a, c^a)$ is updated to $(\mu^a + \delta\mu^a, c^a + \delta c^a)$ with the correction $(\delta\mu^a, \delta c^a)$ computed by solving the following system of linear equation

$$\begin{aligned} K_{ab}^{cc} dc^b + K_{ab}^{c\mu} d\mu^b &= R_a^c + j_x(H)N^a(H) \\ K_{ab}^{\mu c} dc^b + K_{ab}^{\mu\mu} d\mu^b &= R_a^\mu \end{aligned} \quad (35)$$

with the following definitions

$$K_{ab}^{cc} = \int_0^H \left[\frac{1}{\Delta t} N^b N^a\right] dX$$

$$K_{ab}^{c\mu} = -\int_0^H \frac{D}{\beta RT} \frac{\partial N^b}{\partial X} \frac{\partial N^a}{\partial X} dX \quad (36)$$

$$K_{ab}^{\mu c} = -\int_0^H \frac{\partial}{\partial c}\left[ \mu^\Theta + RT \ln \frac{\gamma c}{c_{max} - c} - \frac{\sigma^2}{\rho_{Si}} \frac{\partial}{\partial c}\left(\frac{1}{M}\right) - \frac{2\sigma\beta}{3(1+\beta c)\rho_{Si}} \right] N^b N^a dX$$

$$K_{ab}^{\mu\mu} = \int_0^{h_{0A}} N^b N^a dX$$

The lithium flux $j_X(H)$ applied as boundary condition at anode-electrolyte interface is defined according to the kind of experiment being simulated. For galvanostatic processes the flux is function of the total current intensity $I$ applied to the electrode

$$j_X(\text{H}) = \frac{I_R}{F\rho_{Si}} = \frac{I - I_S}{F\rho_{Si}} \quad (37)$$

while in a potentiostatic test the Butler-Volmer equation (15) returns the total current intensity from the imposed cell voltage $V$

$$j_X(\text{H}) = \frac{I(V) - I_S}{F\rho_{Si}} \quad (38)$$

In both cases the amount of current available to the Faradaic reaction is computed by subtracting the quantity $I_S$ (see eq. (20)) diverted to SEI formation.

The stress measure $\sigma$ can be computed at time $t_{n+1}$ according with

$$\sigma^{n+1} = \sigma^n + \left[ \frac{E}{1-\nu} \frac{\partial}{\partial c}\left(\frac{1-\nu}{E}\right)\sigma^n - \frac{E\beta}{3(1-\nu)(1+\beta c)} \right] \Delta c - \frac{E\dot{\epsilon}_0 \Delta t}{2(1-\nu)} \left(\frac{\sigma^{n+1}}{\sigma_0} - 1\right)^m \quad (39)$$

The material properties $(M, \sigma_0, \dot{\epsilon}_0, m)^n$ are computed from the concentration at the time $t^n$ and equation (39) must be solved iteratively for $\sigma^{n+1}$.

**List of tables**

Table 1: Geometry and operating conditions used in experimental measurements on a-Si electrodes.

| Electrode | Electrode configuration | Substrate thickness $t_s$ (μm) | Residual stress in a-Si film $\sigma_r$ (Gpa) | Cut off potentials (V vs. Li/Li$^+$) | | Number of cycles |
|---|---|---|---|---|---|---|
| | | | | Lithiation | Delithiation | |
| Cell151 | 127 nm a-Si film on 111 wafer | 402 | -0.1 | 0.05 | 0.6 | 4 |
| Cell200 | 103 nm a-Si film on 111 wafer | 400 | -0.36 | 0.05 | 0.6 | 10 |
| PITT | 104 nm a-Si film on 111 wafer | 400 | | 0.05V decrements to 0.05V | 0.05V increments to 1.2V | 1 |

Table 2: Parameters used in modeling thin film electrode. Parameters without cited references were determined through model calibration.

| Parameter | Value |
|---|---|
| Molar density of Si | $7.874 \cdot 10^4$ $mol/m^3$ [59] |
| Mass density of a-Si film | $2.35 \cdot 10^3$ $kg/m^3$ [60] |
| Maximum Li relative molar concentration $c_{max}$ | 3.75 |
| Young modulus of a-Si, $E_0$ | 80 $GPa$ [53] |
| Poisson's ratio of a-Si, $\nu$ | 0.22 [53] |
| Volumetric expansion of Si with Li concentration | $\beta = 0.7$ [61] |
| Faraday constant, F | 96485 C/mol [59] |
| Gas constant, R | 8.314 J K$^{-1}$mol$^{-1}$ [59] |
| Temperature, T | 298 K |
| Factor $M_1$ in Young modulus constitutive law | $-8 \pm 4$ $GPa$ |
| Yield stress of Si at initial Li concentration, $s_0$ | $0.49 \pm 0.08$ $GPa$ |
| Rate of change of flow stress with Li concentration, $s_1$ | $0.07 \pm 0.02$ $GPa$ |
| Characteristic strain rate for plastic flow in Si, $\varepsilon_0$ | $0.64 \cdot 10^{-9} s^{-1}$ |
| Stress exponent for plastic flow, m | 50 |
| Parameter $k_0$ of the Faradaic reaction rate constant | $2.5 \cdot 10^{-8}$ $mol^{1-\alpha} \cdot m^{-2+3\alpha} \cdot s^{-1}$ |
| Parameter $k_1$ of the Faradaic reaction rate constant | $7.5 \cdot 10^{-8}$ $mol^{1-\alpha} \cdot m^{-2+3\alpha} \cdot s^{-1}$ |
| Side reaction rate constant $i_{0,SEI}$ | $1 \cdot 10^{-9}$ $A \cdot m^{-2}$ |
| Equilibrium potential of SEI side reaction $U_{0,SEI}$ | 0.8 V |
| Charge loss in side reactions $Q_{loss}$ | 0.05 C cm$^{-2}$ |
| Stress free rest potential at reference concentration $U_0^{\Theta}$ | 0.74 V |
| $\Omega_2/F$ | 0.8735 V |
| $\Omega_3/F$ | 0.7185 V |
| $\Omega_4/F$ | -4.504 V |
| $\Omega_5/F$ | 6.876 V |
| $\Omega_6/F$ | -4.6272 V |
| $\Omega_7/F$ | 1.1744 V |

# List of figures

Figure 1. Schematic illustration of electrochemical cell and the multi-beam optical sensor setup for curvature measurements, adapted from ref [36]. The inset shows the details of different films grown and deposited on Si wafer. Figure is not to scale.

Figure 2. Idealized model of a Li-ion half-cell consisting of a thin film working electrode bounded to a substrate and Li metal counter-electrode. The Solid Electrolyte Interface layer forms at the anode surface in contact with the electrolyte. SEI is responsible for battery charge loss but it is permeable to Li-ions so it does not significantly reduce lithium intercalation after a stable layer is grown.

Figure 3. (a) Variation of externally applied current $I$ and current associated with lithium insertion reaction $I_R$ with time for cell 151. (b) Measured and predicted voltage plotted vs. time. At the beginning of a charge or discharge cycle the potential undergoes respectively a quick drop or increase. A correct prediction of such fast variation depends on the choice of the rate constants for the Faraday reaction. The overall trend is conditioned by the choice of the self-interaction coefficients of the series expansion representing the excess of free energy due to Li-Li interactions. (c) Variation of stress with time during cycles of charging and discharging.

Figure 4 The plot shows the distribution of the current among main and side reactions during the ten cycles of lithiation and delithiation of Cell 200. The amount of current diverted to SEI formation decreases along with the growth of the SEI layer to a stable thickness.

Figure 5 (a) Comparison of the predicted and measured voltage under cycles of lithiation and delithiation for Cell 200. (b) Comparison of measured and predicted variation of mean stress as a function of capacity during 10 charge-discharge cycles of Cell 200. The green dotted lines emphasize the elastic loading and unloading that characterize the mechanical behavior of the electrode at the beginning of lithiation and delithiation respectively. The stress evolution in plastic flow is well captured through a yield stress constitutive function of lithium concentration.

Fig 6( a) Variation of externally applied voltage with time during the PITT test and the corresponding current density history (b). (c) Stress measured and predicted during the PITT experiment. The stress jump at the beginning of each voltage step is well captured in the simulations. When the voltage reaches values larger than 0.8V the measured stress shows a quick growth observed also in other experiments with Si thin-films delithiated up to 1.2 V.

Figure 7 Comparison of the curve adopted for open circuit potential (by choosing the coefficient of the series expansion of eq.22) and the curves measured in lithiation and delithiation of sputtered amorphous silicon films [42] or obtained with first principle calculations [33].

Figure 8: Values of diffusion coefficient $D$ determined from PITT experiments. The points show values determined using eq. (25) to estimate the apparent diffusion coefficient and correcting the value using (28); the line shows the value determined by fitting full numerical simulations to the PITT data shown in Fig 6(b).

Fig 9 Variations of (a) concentration; (b) plastic strain and (c) stress through the thickness of the film following the voltage step from 0.25 to 0.20V in the PITT experiment. The stress tends to decrease because the yield stress of the electrode decreases with Li concentration.

Fig 10 Variations of (a) concentration; (b) plastic strain and (c) stress through the thickness of the film following the voltage step from 0.30 to 0.35V in the PITT experiment. The stress gradient is higher at the beginning of the voltage step, when Li concentration gradient is also more severe, and it relaxes while the cell voltage is held constant and lithium diffuse through the electrode thickness.

Figure 11 Comparison of predicted and measured current density following a step change in voltage: (a) Transients following a lithiation step; (b) Transients following a de-lithiation step. The current intensity data plotted herein are extracted from the complete set of experimental and numerical results of the PITT experiment represented in Fig. 6b. The prediction of the model is more accurate in the cases considered in Fig 11a, while in delithiation the computation tends to overestimate the slope in correspondence of certain voltage steps.

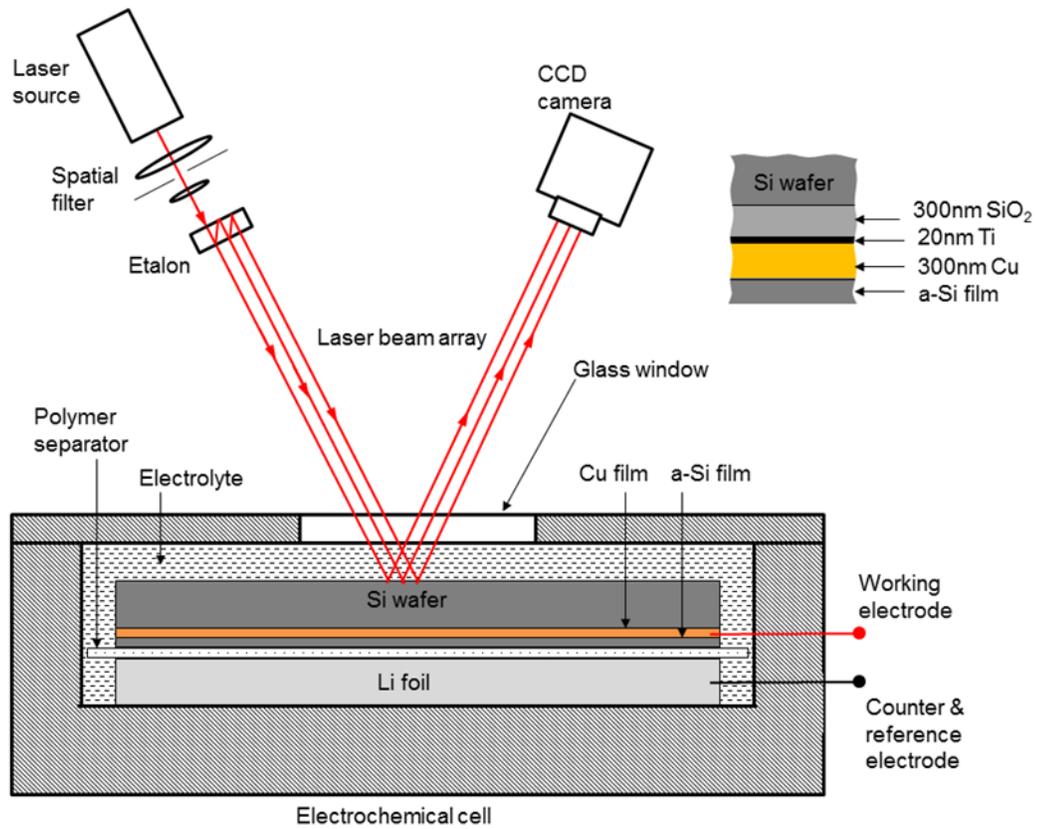

Figure 1. Schematic illustration of electrochemical cell and the multi-beam optical sensor setup for curvature measurements, adapted from ref [36]. The inset shows the details of different films grown and deposited on Si wafer. Figure is not to scale.

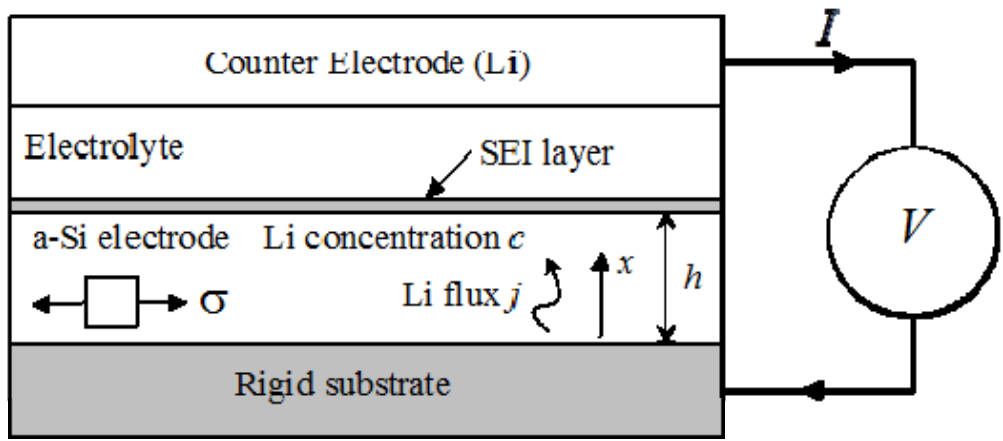

Figure 2. Idealized model of a Li-ion half-cell consisting of a thin film working electrode bounded to a substrate and Li metal counter-electrode. The Solid Electrolyte Interface layer forms at the anode surface in contact with the electrolyte. SEI is responsible for battery charge loss but it is permeable to Li-ions so it does not significantly reduce lithium intercalation after a stable layer is grown.

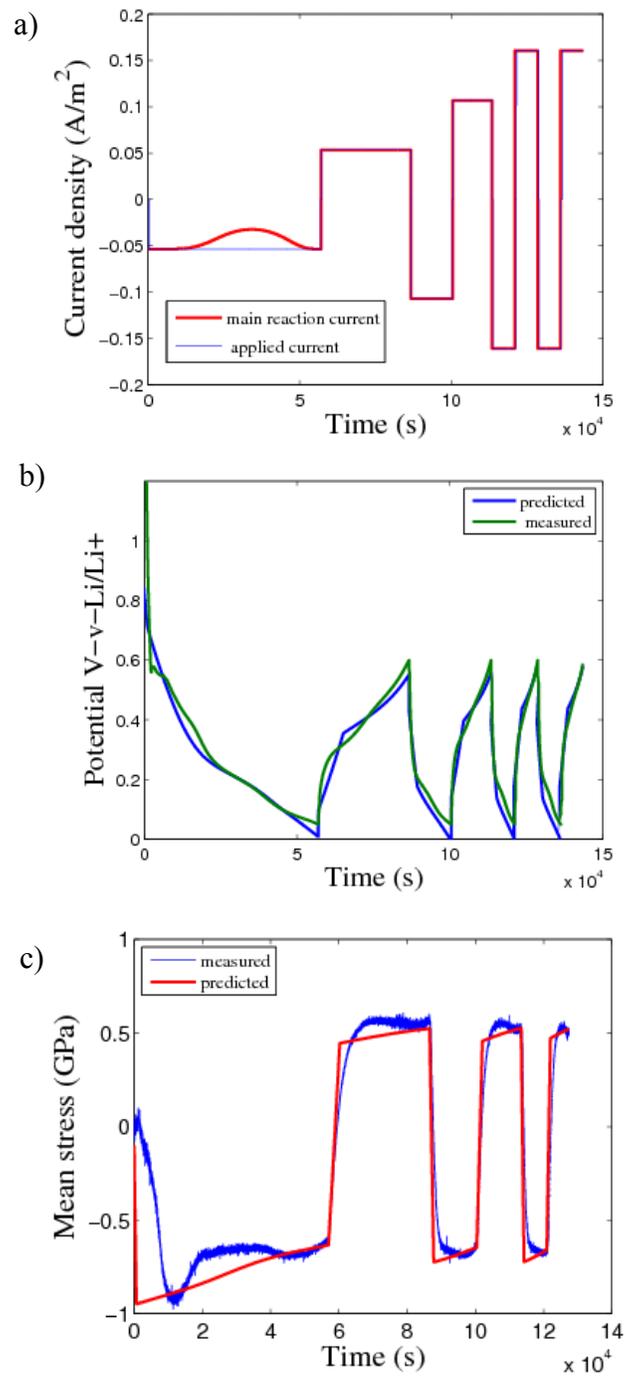

Figure 3. (a) Variation of externally applied current $I$ and current associated with lithium insertion reaction $I_R$ with time for cell 151. (b) Measured and predicted voltage plotted vs. time. At the beginning of a charge or discharge cycle the potential undergoes respectively a quick drop or increase. A correct prediction of such fast variation depends on the choice of the rate constants for the Faraday reaction. The overall trend is conditioned by the choice of the self-interaction coefficients of the series expansion representing the excess of free energy due to Li-Li interactions. (c) Variation of stress with time during cycles of charging and discharging.

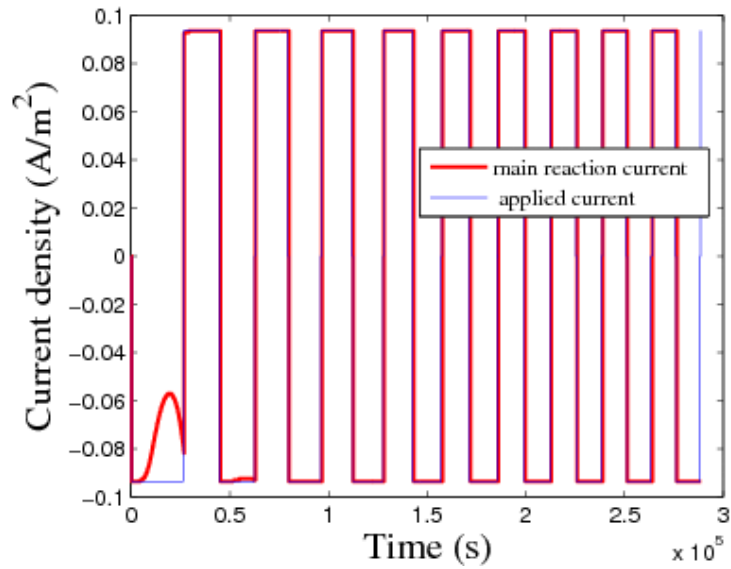

Figure 4 The plot shows the distribution of the current among main and side reactions during the ten cycles of lithiation and delithiation of Cell 200. The amount of current diverted to SEI formation decreases along with the growth of the SEI layer to a stable thickness.

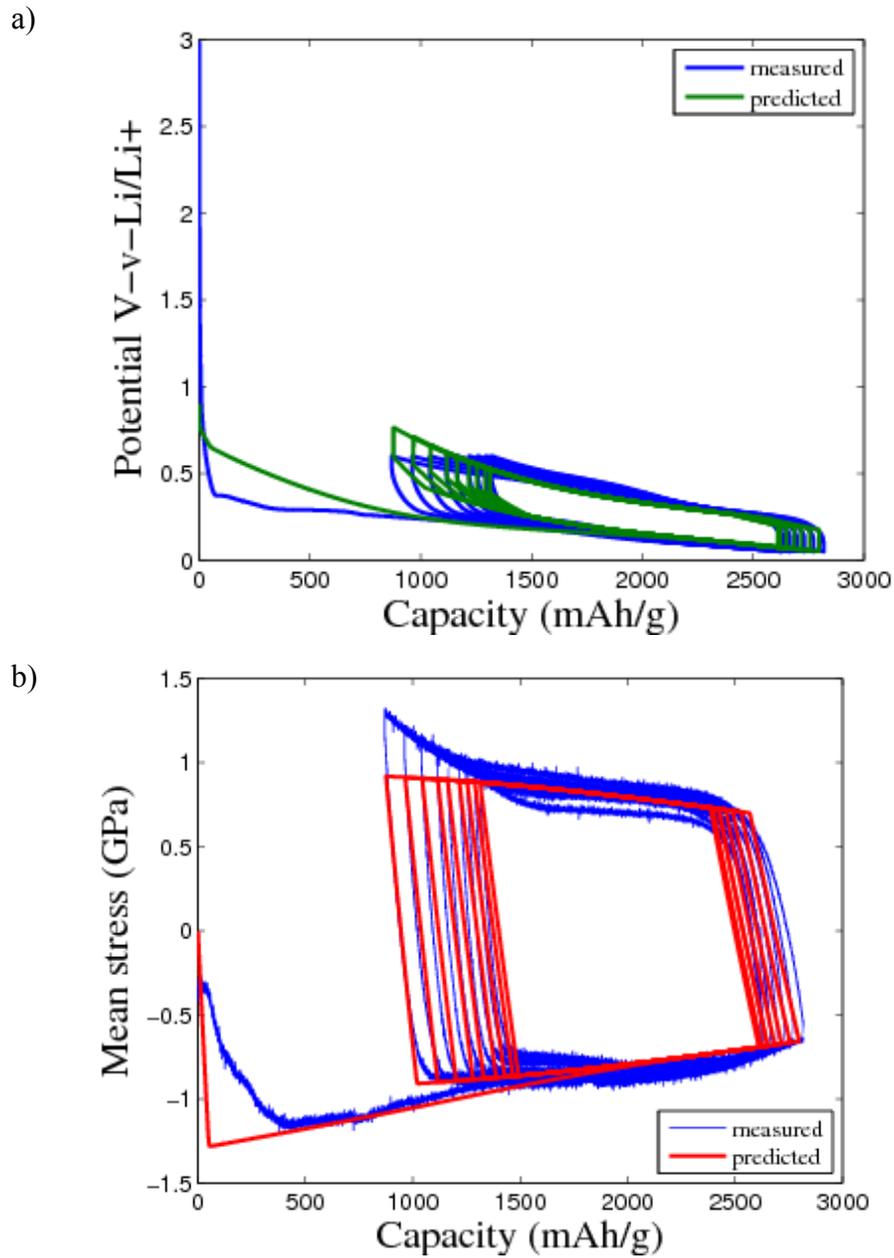

Figure 5 (a) Comparison of the predicted and measured voltage under cycles of lithiation and delithiation for Cell 200. (b) Comparison of measured and predicted variation of mean stress as a function of capacity during 10 charge-discharge cycles of Cell 200. The elastic loading and unloading that characterize the mechanical behavior of the electrode at the beginning of lithiation and delithiation respectively. The stress evolution in plastic flow is well captured through a yield stress constitutive function of lithium concentration.

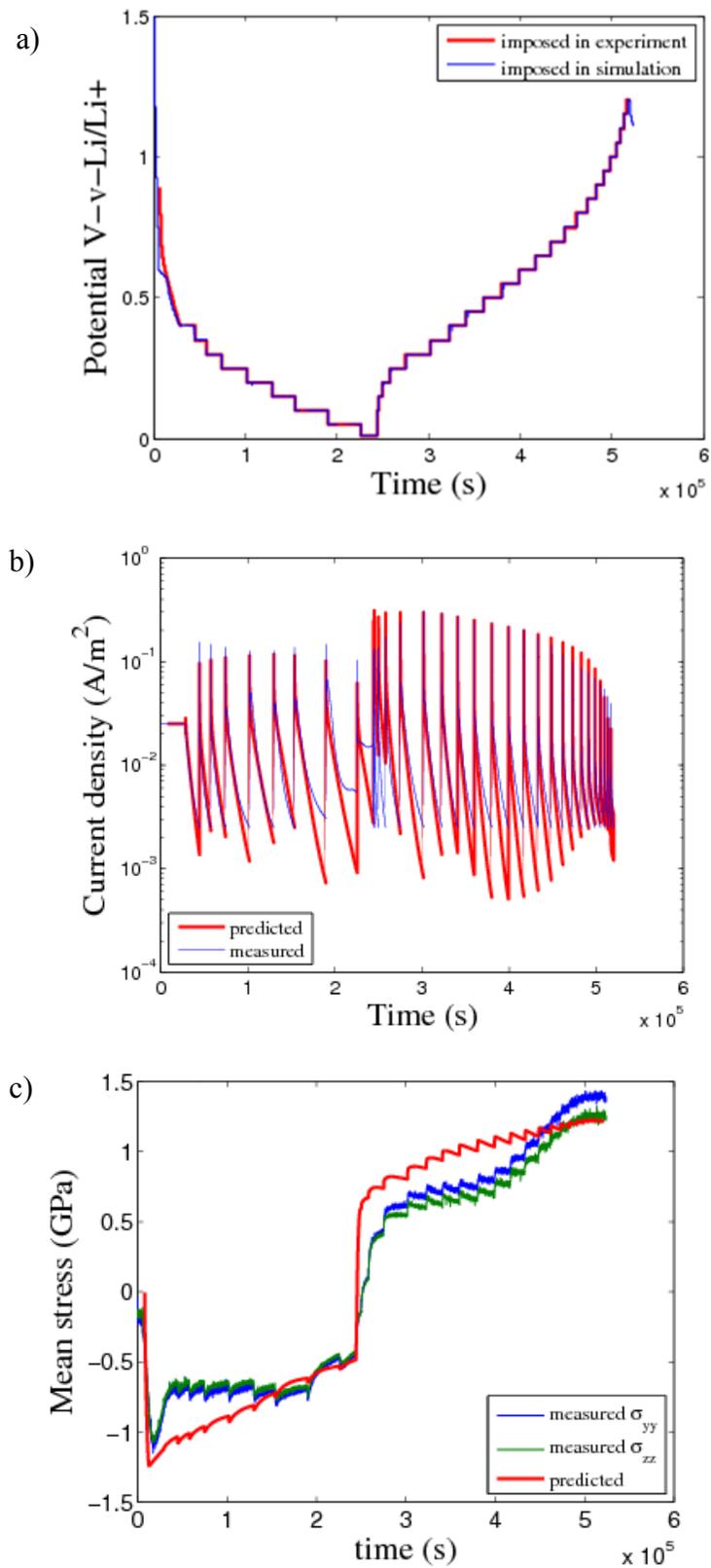

Fig 6( a) Variation of externally applied voltage with time during the PITT test and the corresponding current density history (b). (c) Stress measured and predicted during the PITT experiment. The stress jump at the beginning of each voltage step is well captured in the simulations. When the voltage reaches values larger than 0.8V the measured stress shows a quick growth observed also in other experiments with Si thin-films delithiated up to 1.2 V.

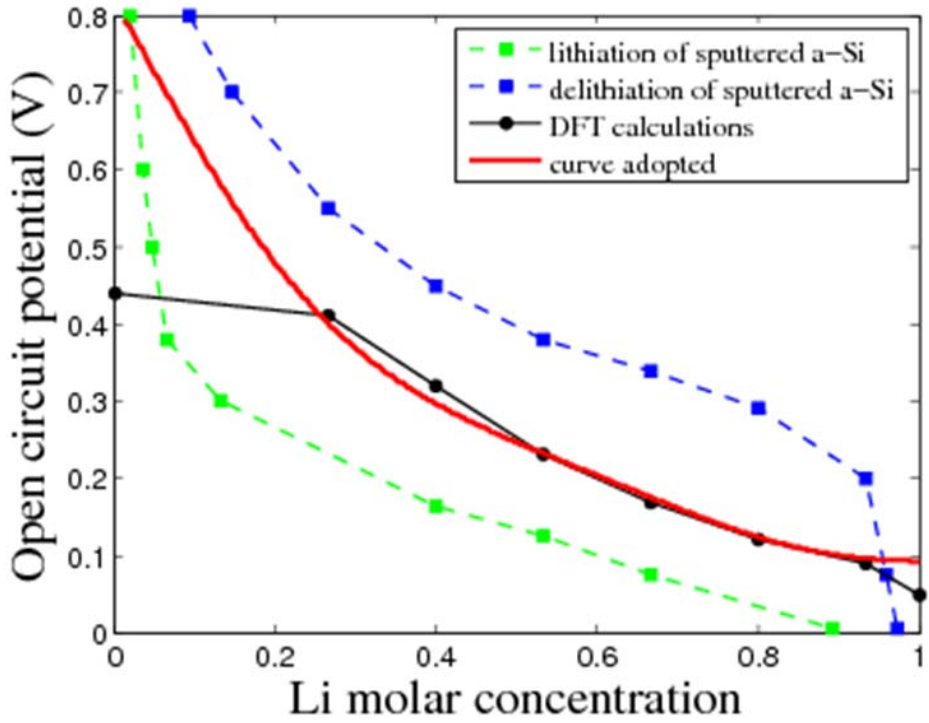

Figure 7 Comparison of the curve adopted for open circuit potential (by choosing the coefficient of the series expansion of eq.22) and the curves measured in lithiation and delithiation of sputtered amorphous silicon films [42] or obtained with first principle calculations [33].

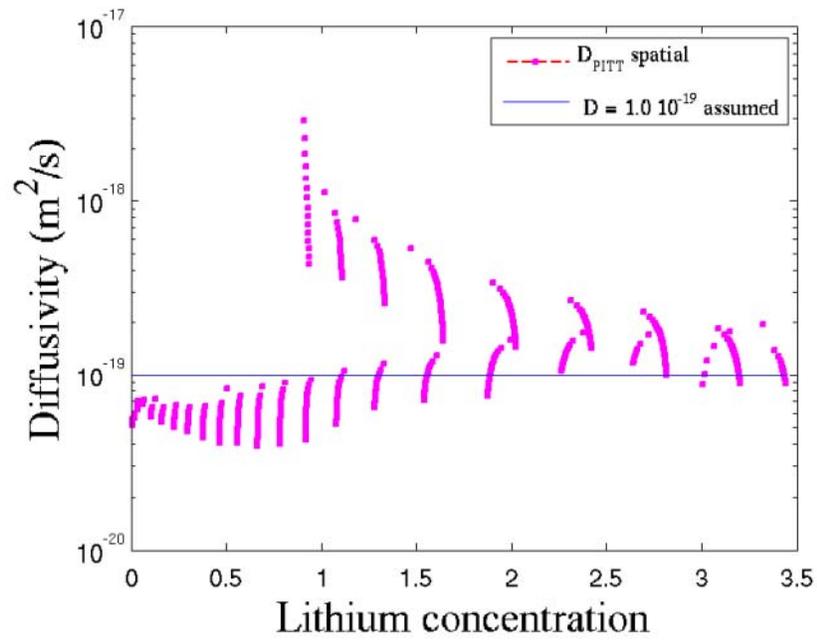

Figure 8: Values of diffusion coefficient $D$ determined from PITT experiments. The points show values determined using eq. (25) to estimate the apparent diffusion coefficient and correcting the value using (28); the line shows the value determined by fitting full numerical simulations to the PITT data shown in Fig 6(b).

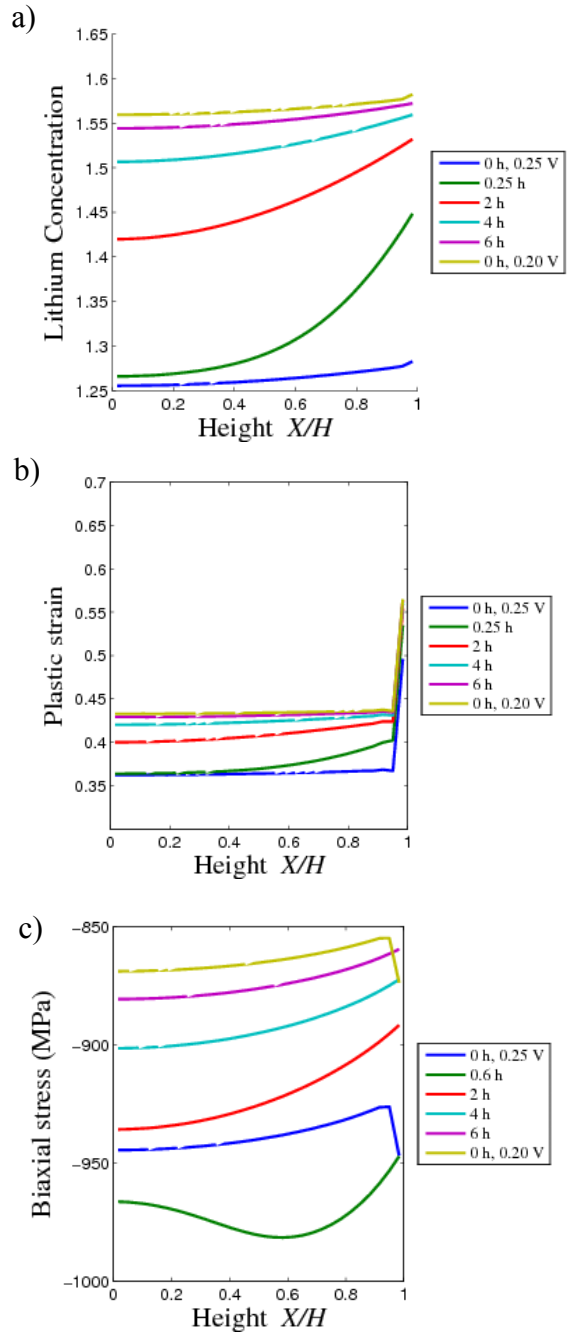

Fig 9 Variations of (a) concentration; (b) plastic strain and (c) stress through the thickness of the film following the voltage step from 0.25 to 0.20V in the PITT experiment. The stress tends to decrease because the yield stress of the electrode decreases with Li concentration.

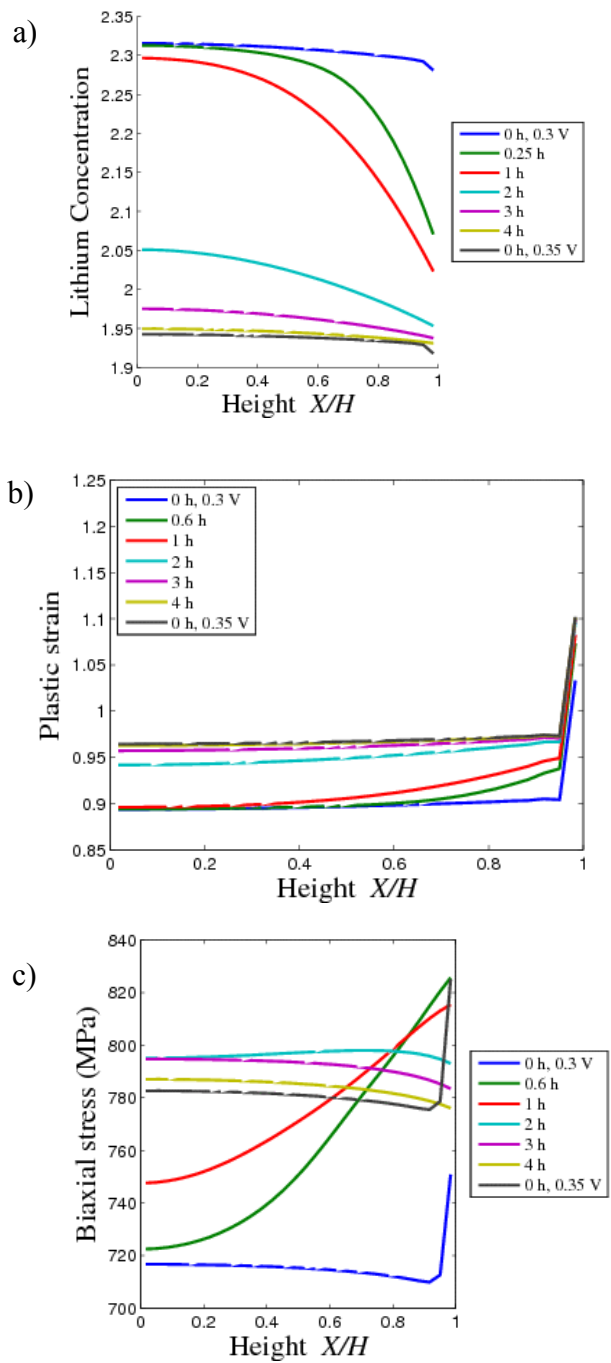

Fig 10 Variations of (a) concentration; (b) plastic strain and (c) stress through the thickness of the film following the voltage step from 0.30 to 0.35V in the PITT experiment. The stress gradient is higher at the beginning of the voltage step, when Li concentration gradient is also more severe, and it relaxes while the cell voltage is held constant and lithium diffuse through the electrode thickness.

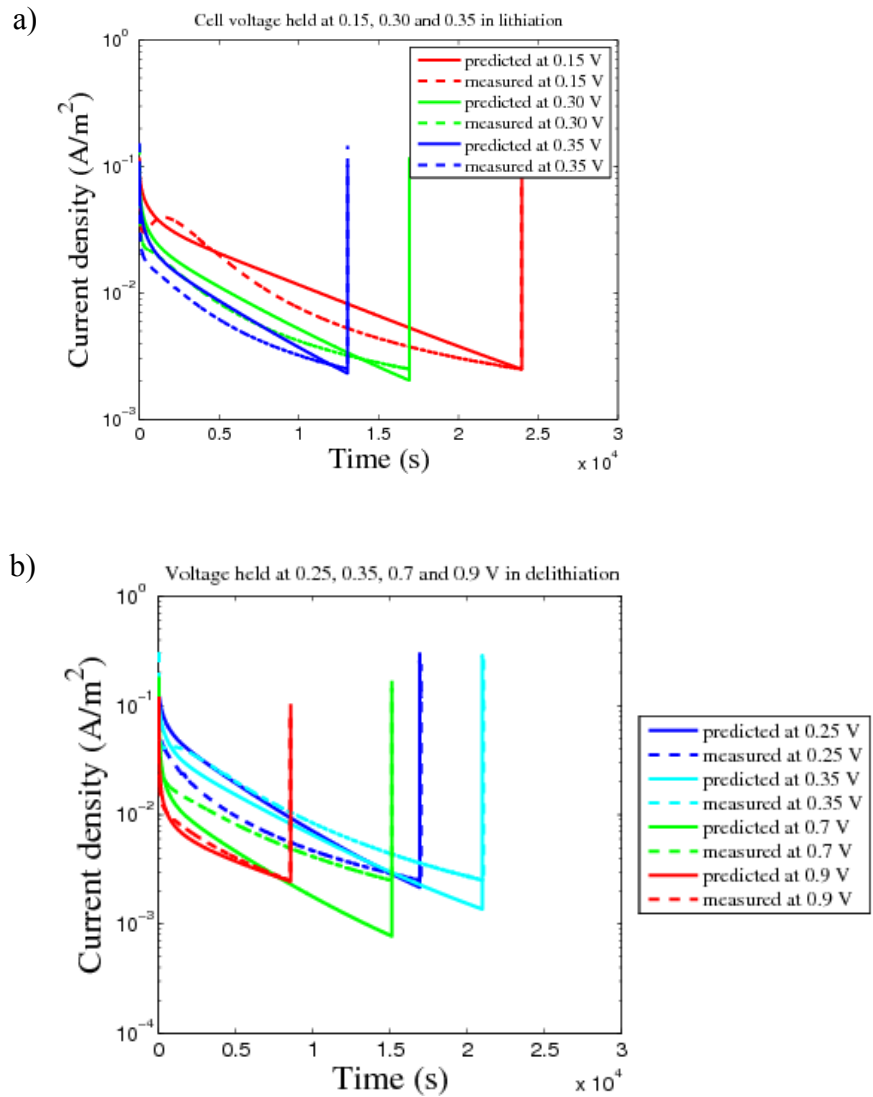

Figure 11 Comparison of predicted and measured current density following a step change in voltage: (a) Transients following a lithiation step; (b) Transients following a de-lithiation step. The current intensity data plotted herein are extracted from the complete set of experimental and numerical results of the PITT experiment represented in Fig. 6b. The prediction of the model is more accurate in the cases considered in Fig 11a, while in delithiation the computation tends to overestimate the slope in correspondence of certain voltage steps.